# About the Mulliken Electronegativity in DFT

*Mihai V. Putz[*], Nino Russo and Emilia Sicilia*

Dipartimento di Chimica and Centro di Calcolo ad Alte Prestazioni per Elaborazioni Parallele e Distribuite-Centro d'Eccellenza MIUR, Università della Calabria, Via Pietro Bucci, Arcavacata di Rende (CS), I-87030, Italy

[*] Permanent address: *Chemistry Department, West University of Timisoara, Str. Pestalozzi No.16, Timisoara, RO-300115, Romania.*

**Abstract**

In the framework of density functional theory a new formulation of electronegativity that recovers the Mulliken definition is proposed and its reliability is checked by computing electronegativity values for a large number of elements. It is found that the obtained values, which are compared with previously proposed electronegativity scales, fulfill the main periodic criteria.

**Introduction**

In modern chemistry, the concept of electronegativity ($\chi$), firstly proposed by J. J. Berzelius in 1811, plays a crucial role because it can be considered as one of the most important chemical descriptors in order to account for the tendency of the atoms to build up a molecular system.

About 70 years ago, Pauling by an ingenious mixing of thermodynamical and quantum mechanical arguments introduced an electronegativity scale, which has enriched the concept of atomic periodic properties [1].

A step forward was took in 1934 and 1935 [2] by Mulliken, which introduced a different formulation in terms of two other periodic properties, namely the ionization potential and the electron affinity, and enabled the extension of this concept to molecules.

In the history of electronegativity formulations has to be remarked the classical Allred-Rochow scale that introduces the idea of force into the electronegativity theory [3].

The electronegativity definition was subsequently modified and enriched toward a gradual appreciation of the various complexities involved in the concept [4-21].

In a continuous effort to better define the rather intuitive concept of electronegativity, the Parr idea to define $\chi$ as the negative of the chemical potential of the density functional theory (DFT) of Hohenberg and Kohn, i.e. as the first derivative of the energy functional, the connection between electronegativity and quantum mechanics has been established [21, 22]. This result has opened up exciting perspectives to calculate $\chi$ for whatever many electron systems such as atoms, molecules, clusters. Moreover, in order to measure the "power of atoms to attract electrons to themselves" [1], using the arsenal of different formulation, working formulas and modern quantum mechanical methods, a series of electronegativity scales has been proposed [23].

In this work, a new electronegativity formulation is proposed within the density functional theory, which appears to be the natural and convenient tool to investigate this quantity. In the same context some attention is given to the analysis of the equivalence between the Mulliken and the differential definitions of $\chi$. The calculated electronegativities for 52 atoms are organized in an electronegativity scale that is compared with those previously proposed [24-26] using several definitions. Furthermore, the orbital electronegativities for C, N, and O atoms are also given.

**Theoretical Method**

*Density Functional reactivity indices background*

Among the chemical concepts that have found a rigorous quantitative definition in the framework of the DFT a special attention was given to the electronegativity formulation [27, 28].

For an *N*-electronic system placed into an external potential $V(r)$ the general (first order) equation of the change in the chemical potential, $\mu = \mu[N, V(r)]$, can be written as [21]:

$$d\mu = 2\eta dN + \int f(r) dV(r) dr \qquad (1)$$

in which the variation of the chemical potential $\mu$ (or the electronegativity in the Parr definition $\mu = -\chi$) for an electronic system is correlated with variation of the number of electrons and of the external potential through the chemical hardness ($\eta$):

$$2\eta = \left(\frac{\partial \mu}{\partial N}\right)_{V(r)} \qquad (2)$$

and the Fukui function (*f*):

$$f(r) \equiv \left(\frac{\delta \mu}{\delta V(r)}\right)_N \qquad (3)$$

Thus, the chemical potential (or the electronegativity) concept appears to be strictly connected with the other two, chemical hardness and Fukui function, extensively used reactivity criteria. It is noteworthy that although in the original hardness definition the factor ½ was put in to make it symmetrical with respect to the chemical potential definition [29], nowadays the convention without this factor is also used [30].

In a similar way (see equation 1) the total energy for the electronic system, $E = E[N, V(r)]$, can be written as:

$$dE = \mu dN + \int \rho(r)dV(r)dr \tag{4}$$

where the chemical potential and the electronic density, $\rho(r)$, are defined as:

$$\mu = \left(\frac{\partial E}{\partial N}\right)_{V(r)} \tag{5}$$

$$\rho(r) = \left(\frac{\delta E}{\delta V(r)}\right)_N . \tag{6}$$

The equation 4 can be rewritten in terms of the Maxwell identities as:

$$\left(\frac{\delta \mu}{\delta V(r)}\right)_N = \left(\frac{\partial \rho(r)}{\partial N}\right)_{V(r)} . \tag{7}$$

Using the Parr definition of electronegativity for the chemical potential, from equation 5 this chemical descriptor takes the form:

$$\chi(N) = -\left(\frac{\partial E}{\partial N}\right)_{V(r)} . \tag{8}$$

Furthermore, using the same equation 5 incorporated in the hardness definition, equation 2, the expression for the hardness, as the second order derivative of the total energy with respect to the total number of electrons, assumes the form:

$$\eta(N) = \frac{1}{2}\left(\frac{\partial^2 E}{\partial N^2}\right)_{V(r)} . \tag{9}$$

Turning now to the Maxwell identity, equation 7, the Fukui index given by equation 3 can be defined in terms of the density and the number of electrons as:

$$f(r) = \left(\frac{\partial \rho(r)}{\partial N}\right)_{V(r)} \tag{10}$$

Combining in the equation 1, the expressions 2 and 10, the following differential equation for the electronegativity is obtained:

$$d\chi = \left(\frac{\partial \chi}{\partial N}\right)_{V(r)} dN - \int \left(\frac{\partial \rho(r)}{\partial N}\right)_{V(r)} dV(r)dr \tag{11}$$

Taking into account the relation 7 expressed within the Parr electronegativity definition,

$$\left(\frac{\partial \rho(r)}{\partial N}\right)_{V(r)} = -\left(\frac{\delta \chi}{\delta V(r)}\right)_N \tag{12}$$

it is easily recognized that equation 11 has the same form as equations 1 and 4.

However, in order to find the electronegativity we propose the alternative integration of equation 1 in the following way.

First, let us express the hardness and Fukui function through the relations [21]:

$$2\eta = \frac{1}{S} \tag{13}$$

$$f(r) = \frac{s(r)}{S} \tag{14}$$

where $S$ and $s(r)$ represent the global and the local softness defined as:

$$S = \left(\frac{\partial N}{\partial \mu}\right)_{V(r)} \tag{15}$$

$$s(r) = \left(\frac{\partial \rho(r)}{\partial \mu}\right)_{V(r)}. \tag{16}$$

Assuming that

$$N = \int_{-\infty}^{+\infty} \rho(r) dr \tag{17}$$

the connection between the global and the local softness indices can be obtained:

$$S = \int_{-\infty}^{+\infty} s(r) dr \tag{18}$$

Applying to the exact formula of Berkowitz and Parr [31a] and Ayers [31b], relating the conventional linear response function $[\delta \rho(r)/\delta V(r')]_N$ with softness, local softness and softness kernel, the three quantum mechanical constraints such as the translational invariance condition [31c], the Hellmann-Feynman theorem [31d], and the normalization of the linear response function [31e], the softness kernel $s(r, r')$ becomes [31e]:

$$s(r,r') = L(r')\delta(r-r') + \rho(r)\rho(r') \tag{19}$$

as a sum of local and non-local contributions. Now, the local response function:

$$L(r) = -\frac{\nabla \rho(r)}{\nabla V(r)} \qquad (20)$$

corresponds to the scalar quantity $L(r) = -\nabla \rho(r) \cdot \nabla V(r) / |\nabla V(r)|^2$.

This model is general beside the different ways to evaluate the non-local term in equation 19.

Integrating the equation 19 over $r'$ the relation 16 for the local softness becomes:

$$s(r) = L(r) + N\rho(r) \qquad (21)$$

and using the relations 17 and 18 the corresponding global softness looks like:

$$S = \int_{-\infty}^{+\infty} L(r)dr + N^2 \qquad (22)$$

All these chemical descriptors depend on the electronic density and will be usefully combined in order to derive the desired Mulliken electronegativity.

*The Absolute Electronegativity Formula*

Introducing the expressions 21 and 22 into the hardness and Fukui functions definitions (see equations 13 and 14) we can integrate the equation 1 for the electronegativity, assuming the initial zero electronegativity value as $V(x) \to 0$, to obtain:

$$\chi(N) = -\int_0^N 2\eta dN - \int_{-\infty}^{+\infty} f(r)V(r)dr$$

$$= -\int_0^N \frac{1}{S} dN - \frac{1}{S}\int_{-\infty}^{+\infty} s(r)V(r)dr$$

$$= -\int_0^N \frac{dN}{\int_{-\infty}^{+\infty} L(r)dr + N^2} - \frac{\int_{-\infty}^{+\infty} L(r)V(r)dr + N\int_{-\infty}^{+\infty} \rho(r)V(r)dr}{\int_{-\infty}^{+\infty} L(r)dr + N^2} . \qquad (23)$$

In order to simplify the expression 23, the following definitions are introduced:

$$a = \int_{-\infty}^{+\infty} L(r)dr = \int_{-\infty}^{+\infty} \frac{\nabla\rho(r)}{[-\nabla V(r)]}dr \quad (24)$$

$$b = \int_{-\infty}^{+\infty} L(r)V(r)dr = \int_{-\infty}^{+\infty} \frac{\nabla\rho(r)}{[-\nabla V(r)]}V(r)dr \quad (25)$$

$$C_A = \int_{-\infty}^{+\infty} \rho(r)V(r)dr \quad . \quad (26)$$

The last introduced quantity, $C_A$, has been already named *chemical action* index [32] since, analogously to the physical definition of an "action", will be shown that the variational principle can be applied to it.

The integration of the equation 23 gives the final and general expression for the electronegativity:

$$\chi(N) = -\frac{1}{\sqrt{a}}\arctan[\frac{N}{\sqrt{a}}] - \frac{b}{a+N^2} - NC_A\frac{1}{a+N^2} \quad . \quad (27)$$

which is an analytical counterpart of a more general dependence of $\chi$ on the energy functional. This correspondence can be obtained employing in the general energy equation 4 the ground state constrain of constant chemical potential. The equation of the change of the total energy (equation 4) can be rewritten as:

$$dE - \mu dN = \int \rho(r)dV(r)dr \quad . \quad (28)$$

which corresponds to a path integral over the reaction path followed by the functional differentiation around the ground state:

$$\delta\{\int [dE - \mu dN]\} = \delta\{\int [\int \rho(r)dV(r)dr]\} \quad . \quad (29)$$

Under the ground state constrain [18d, 20]:

$$\mu = CONSTANT \quad (30)$$

it is provided the equivalence:

$$\delta\{E[\rho] - \mu N[\rho]\} = \delta C_A \quad (31)$$

that appears to be the most general relationship between chemical potential (negative of electronegativity) and the total energy, through the chemical action.

*The Mulliken Electronegativity from DFT Principles*

Starting from the proposed electronegativity expression (eq. 27), the identification of the Mulliken and the Parr differential electronegativity definitions is demonstrated, without to have recourse the application of the finite difference scheme to equation 8. Moreover, a reformulation of the Mulliken electronegativity definition is obtained as a generalization of the classical one when the DFT concepts are included.

Starting from the traditional Mulliken electronegativity formula, in terms of ionization potential (*IP*) and electron affinity (*EA*) [2], the following series of identities can be considered:

$$\chi_M(N) = \frac{IP(N) + EA(N)}{2} \equiv \frac{[E(N) - E(N+1)] + [E(N-1) - E(N)]}{2} = \frac{-E(N+1) + E(N-1)}{2} \quad (32)$$

The corresponding integral formulation of the last member of eq. 32 looks like:

$$\chi_M(N) = -\frac{1}{2} \int_{|N-1\rangle}^{|N+1\rangle} dE_N \quad (33)$$

If the eq. 4 is introduced into eq. 33 and the integration limits are taken into account, results:

$$\chi_M(N) = -\frac{1}{2} \int_{|N-1\rangle}^{|N+1\rangle} \left[ \left(\frac{\partial E}{\partial N}\right)_{V(r)} dN + \int \rho_N(r) dV_N(r) dr \right]$$

$$= -\frac{1}{2} \int_{N-1}^{N+1} \left(\frac{\partial E}{\partial N}\right)_{V(r)} dN - \frac{1}{2} \left[ \int \rho_N(r) V_{N+1}(r) dr - \int \rho_N(r) V_{N-1}(r) dr \right] . \quad (34)$$

As a consequence of the Hohenberg-Kohn theorems, each of the last two terms of the right side of equation 34 vanish. This can be accomplished in the virtue of the equivalence 31, that permits to rewrite the variational equation for the ground-state density functional theory:

$$\delta\{E[\rho] - \mu N[\rho]\} = 0 \quad (35)$$

as:

$$\delta C_A = \delta\left[\int \rho(r) V(r) dr\right] = 0 \quad (36)$$

It is worth to note that the last expression combines the first and second Hohenberg-Kohn theorems, providing the context in which the last two terms of eq. 34 become zero. Thus, the Mulliken electronegativity within *DFT* is obtained:

$$\chi_M(N) = -\frac{1}{2}\int_{N-1}^{N+1}\left(\frac{\partial E}{\partial N}\right)_{V(r)} dN \tag{37}$$

As eq. 37 clearly shows, the relationship between the Mulliken electronegativity and the Parr differential one does not involve the use of the finite difference approximation and does not depend on the particular form of *E(N)*.

Taking into account equation 8, the identity in eq. 37, becomes:

$$\chi_M(N) = \frac{1}{2}\int_{N-1}^{N+1}\chi(N)dN \;. \tag{38}$$

The result 38, which is here rigorously density functional principles based, was also previously proposed by Komorowski [33], considering the average of the Parr differential electronegativity over a suitable region of charge.

By performing the definite integration required in the equation 38, using equation 27, we arrive to the present density functional Mulliken version of electronegativity:

$$\chi_M(N) = \frac{b+N-1}{2\sqrt{a}}\arctan\left(\frac{N-1}{\sqrt{a}}\right) - \frac{b+N+1}{2\sqrt{a}}\arctan\left(\frac{N+1}{\sqrt{a}}\right) + \frac{C_A - 1}{4}\ln\left[\frac{a+(N-1)^2}{a+(N+1)^2}\right]. \tag{39}$$

Although we have started from a formulation in terms of ionization and affinity terms we have derived a reformulation depending on different quantities, such as total number of electrons, density, external potential and their gradients. This new approach allows to extend the description of various chemical situations as atoms involved in bonds.

Equations 27 and 39 are consistent with the electronegativity equalization principle as shown in the Appendix. The electronegativity expression given in the equation 39 represents our proposed working electronegativity formula that will be used in order to derive an electronegativity scale for almost the entire periodic table.

**Computational Details**

In this section we list the steps that have to take into account in order to apply the proposed Mulliken electronegativity formula.

The electrons of the atomic system are distinguished as core and valence ones.

The core produces an effective potential in which the valence electrons are moving.

The effective potential of the core is represented as a pseudo-potential. In this work, the Stuttgart/Bonn group pseudo-potentials have been employed, starting from lithium [34].

The valence electrons are represented by the pseudo-potential wave-function [34,] which is forced normalize to one, to simplify the computations. To fulfill the normalization condition:

$$\int |\psi(r)|^2 dr = 1 \qquad (40)$$

a sort of "scaling factor" ($q$) has to be involved in the wave function expansion:

$$\psi = \sum_i A_i \exp(-q\alpha_i r^2) \qquad (41)$$

and the Mulliken electronegativity (39) is computed.

Since the condition of eq. 40 is very restrictive, the scaling factor, $q$, has been fixed also considering an additional constrain that takes into consideration the number of valence electrons. This aim was accomplished noting that previous atomic electronegativity formulations fit our definition of chemical action, $C_A$, through the use of a Coulombic potential [35]:

$$\chi(N,Z) = \left\langle \frac{1}{r} \right\rangle = \int \left\{ \rho(N,Z,r) \frac{1}{r} \right\} dr = -\int \left\{ \rho(N,Z,r) V_{COULOMB}(r) \right\} dr \equiv -C_A^{COULOMB} \qquad (42)$$

with $Z$ equal to the nuclear charge.

**Results and Discussion**

Before to discuss the reliability of our results, we would like to underline that the main goal of our work resides in the demonstration that the Mulliken electronegativity can be rigorously expressed within the density functional theory and, consequently, we do not propose a new scale that aspires to replace the previous electronegativity scales. However, we emphasize that although the starting point for the present development is the wide accepted approximate Mulliken formulation we have reformulated it in terms of quantities that better take into account the whole information of an atomic system.

**Table 1.** The atomic Mulliken electronegativities (in eV) computed at different levels of theory and experiment, from top to bottom of each element's cell: present, experimental [25, 37], Mulliken-Jaffé [24, 36], and Xα [26] electronegativities, respectively.

| Li | Be | | | | | | | | | | | B | C | N | O | F | Ne |
|---|---|---|---|---|---|---|---|---|---|---|---|---|---|---|---|---|---|
| 3.02 | 3.40 | | | | | | | | | | | 5.66 | 8.58 | 9.77 | 12.41 | 15.60 | 13.37 |
| 3.01 | 4.9 | | | | | | | | | | | 4.29 | 6.27 | 7.27 | 7.53 | 10.41 | - |
| 3.10 | 4.80 | | | | | | | | | | | 5.99 | 7.98 | 11.5 | 15.25 | 12.18 | 13.29 |
| 2.58 | 3.8 | | | | | | | | | | | 3.4 | 5.13 | 6.97 | 8.92 | 11.0 | 10.31 |
| **Na** | **Mg** | | | | | | | | | | | **Al** | **Si** | **P** | **S** | **Cl** | **Ar** |
| 2.64 | 3.93 | | | | | | | | | | | 5.89 | 6.80 | 8.33 | 11.88 | 14.06 | 12.55 |
| 2.85 | 3.75 | | | | | | | | | | | 3.21 | 4.76 | 5.62 | 6.22 | 8.30 | - |
| 2.80 | 4.09 | | | | | | | | | | | 5.47 | 7.30 | 8.90 | 10.14 | 9.38 | 9.87 |
| 2.32 | 3.04 | | | | | | | | | | | 2.25 | 3.60 | 5.01 | 6.52 | 8.11 | 7.11 |
| **K** | **Ca** | **Sc** | **Ti** | **V** | **Cr** | **Mn** | **Fe** | **Co** | **Ni** | **Cu** | **Zn** | **Ga** | **Ge** | **As** | **Se** | **Br** | **Kr** |
| 2.48 | 2.19 | 1.83 | 2.28 | 2.42 | 2.72 | 2.01 | 3.90 | 3.03 | 3.48 | 2.91 | 3.13 | 3.30 | 4.24 | 4.94 | 4.82 | 7.35 | 9.59 |
| 2.42 | 2.2 | 3.34 | 3.45 | 3.6 | 3.72 | 3.72 | 4.06 | 4.3 | 4.40 | 4.48 | 4.45 | 3.2 | 4.6 | 5.3 | 5.89 | 7.59 | - |
| 2.90 | 3.30 | 4.66 | 5.2 | 5.47 | 5.56 | 5.23 | 6.06 | 6.21 | 6.30 | 4.31 | 4.71 | 6.02 | 8.07 | 8.3 | 9.76 | 8.40 | 8.86 |
| 1.92 | 1.86 | 2.52 | 3.05 | 3.33 | 3.45 | 4.33 | 4.71 | 3.76 | 3.86 | 3.95 | 3.66 | 2.11 | 3.37 | 4.63 | 5.91 | 7.24 | 6.18 |
| **Rb** | **Sr** | **Y** | **Zr** | **Nb** | **Mo** | **Tc** | **Ru** | **Rh** | **Pd** | **Ag** | **Cd** | **In** | **Sn** | **Sb** | **Te** | **I** | **Xe** |
| 1.05 | 1.63 | 1.76 | 1.73 | 1.68 | 2.07 | 1.96 | 1.93 | 1.72 | 1.98 | 2.18 | 2.36 | 2.48 | 2.74 | 6.29 | 4.98 | 6.70 | 6.27 |
| 2.34 | 2.0 | 3.19 | 3.64 | 4.0 | 3.9 | - | 4.5 | 4.3 | 4.45 | 4.44 | 4.43 | 3.1 | 4.30 | 4.85 | 5.49 | 6.76 | - |
| 2.09 | 3.14 | 4.25 | 4.57 | 5.38 | 7.04 | 6.27 | 7.16 | 7.4 | 7.16 | 6.36 | 5.64 | 5.28 | 7.90 | 8.48 | 9.66 | 8.10 | 7.76 |
| 1.79 | 1.75 | 2.25 | 3.01 | 3.26 | 3.34 | 4.58 | 3.45 | 3.49 | 3.52 | 3.55 | 3.35 | 2.09 | 3.20 | 4.27 | 5.35 | 6.45 | 5.36 |

In any case the obtained results for 52 elements from Lithium through Xenon seem to satisfy most of the criteria for the acceptability of an electronegativity scale.

Results coming from the application of equation 39 are listed in Table 1 together with some previous electronegativity scales.

In particular, we have reported the Mulliken electronegativity, named experimental, obtained using experimental values of ionisation potentials and electron affinities [25, 37].

Amongst the theoretical approaches, we have chosen to report the Mulliken-Jaffé scale [24, 36] and the electronegativity values, calculated by a simple $X\alpha$ method employing the transition-state approach [26], that we have called $X\alpha$ scale.

Concerning the acceptability guidelines our results can be summarized as follows [38]:

(i) the scale was built up for isolated atoms;

(ii) two significant figures are able to distinguish the electronegativities of all the considered elements;

(iii) the given values are expressed in electronvolts;

(iv) all the valence electrons were included in the electronegativity computation;

(v) values obtained for the elements (N, O, F, Ne, Cl, Ar, Br, Kr) that present oxidation states lower than their valence electrons follow the appropriate requirement with the exception of the chlorine and fluorine atoms that have a $\chi$ values higher than that of the nearest noble gas atoms, Ne and Ar, respectively. The electronegativity trend for these atoms is F>Cl>Ne>Ar>O>N>Kr>Br. We consider that the use of functions of spherical symmetry can be indicated as the main source of error in the chlorine and fluorine electronegativity determinations;

(vi) the six considered metalloid elements (B, Si, Ge, As, Sb, Te) that separate the metals from the non-metals have electronegativity values, which do not allow overlaps between metals and non-metals. Furthermore, looking at the $\chi$ metal values the requirement that they have to have electronegativities lower than that found for the silicon, the so-called silicon rule [38], is satisfied although silicon does not show the lowest $\chi$ value in the metalloid band. On the other hand this result well agrees with the experimental [37] and previous theoretical [26] determination performed at $X\alpha$ level of theory. Finally we note that this behavior respects the fundamental rule of the decreasing of electronegativity down groups;

(vii) for binary compounds the difference in electronegativity satisfies the definition of ionic, covalent and metallic bonds as required by the Ketelaar's triangle;

(viii) our definition is fully quantum mechanical;

(ix) the decreasing of $\chi$ along the group is respected as well as its difference in going from light to heavy atoms of the same period increases left to right across rows taking into account that for some heavy elements the relativistic effects, which are not considered in the computations, can affect this trend. Correctly, the halogen atoms have the highest electronegativity values with respect to their left row neighbors. Looking to the transition metal atoms, we underline that the obtained electronegativities fall in a narrow range of values compared with those of the main group atoms.

In order to verify the adopted symmetry influence, we have recalculated the electronegativity for C, O and N atoms by using p-type orbitals and the sp, $sp^2$ and $sp^3$ hybridization states.

Results, reported in Figure 1, show how the introduction of more realistic basis sets for the valence orbital description increases the agreement amongst electronegativity trends.

Indeed, the computed $\chi$ values follow the trends previously obtained by Mulliken – Jaffè by using s- and p- orbital basis and the same hybridized states.

Finally, it should be remarked analysing Fig.1, that the electronegativity trend from a type of hybridization to another is similar.

This indicates that the actual electronegativity formulation preserves also the orbital hierarchy and is sensitive to the hybrid orbitals as well.

Moreover, the usefulness of this approach can be tested performing calculation of reactivity indices and of other periodic properties [40].

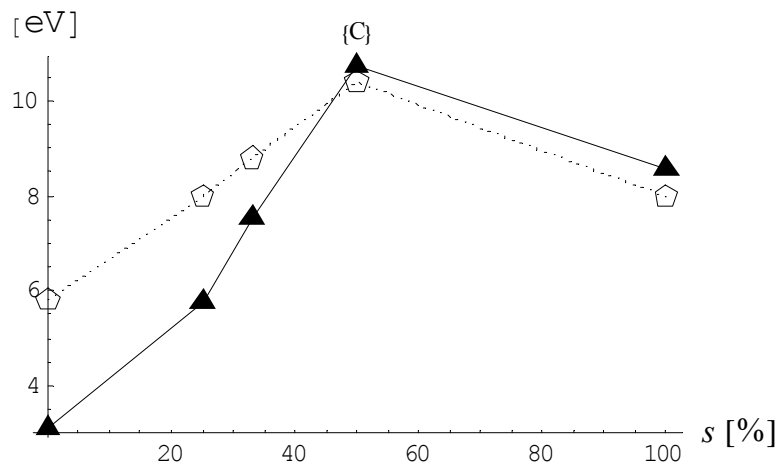
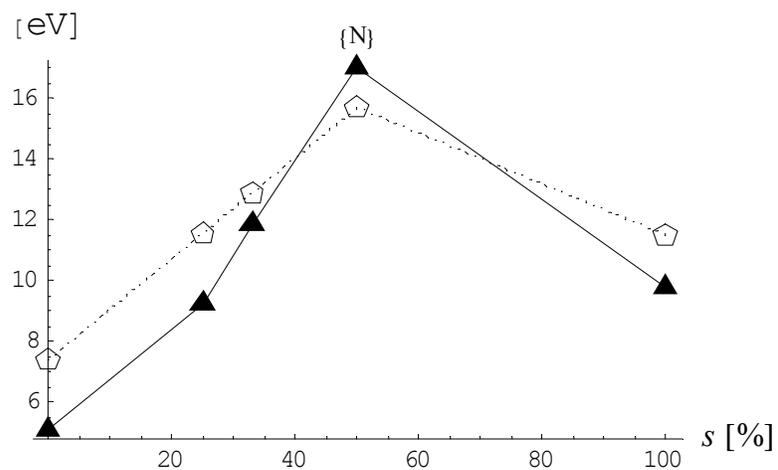
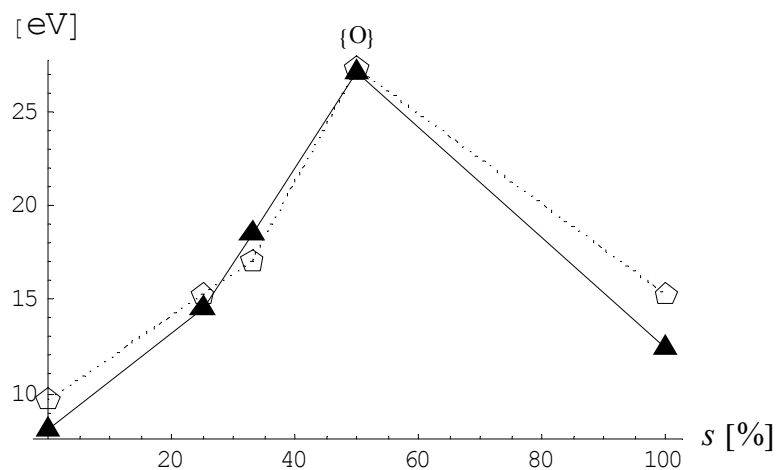

**Figure 1.** Orbital electronegativities $\chi$ for C, N and O atoms, from top to bottom respectively, versus the percent contribution of s orbital obtained by using the basis set (BS) methods together with electronegativity values from Mulliken-Jaffè (MJ) scale [39]. The used symbols are interpreted as: ▲ for $\chi^{BS}$, and ⬠ for $\chi^{MJ}$ in each plot, respectively.

**Conclusions**

In the framework of density functional theory, we propose a new electronegativity formulation which is used to compute a scale for almost the entire periodic table. Furthermore, we show an analytical way for the determination of the density functional Mulliken electronegativity that properly links the electronic local quantities, such as the density and the potential, with the global ones, as the number of electrons neither involving the direct energy computation nor assuming a particular behavior for energy, $E(N)$.

The proposed electronegativity values follow almost all the general criteria of acceptability of the proposed $\chi$ scales, although the data are basis-set dependent on the way in which the valence orbital electrons are described. The present approach opens also other perspectives. For instance, the actual study can be extended also to the molecular systems by the appropriate implementation of the molecular density and the (effective) potential.

**Acknowledgments**

The authors gratefully acknowledge Università della Calabria and MIUR for the financial support. Permanent address of MVP: *Chemistry Department, West University of Timisoara, Str. Pestalozzi No.16, Timisoara, RO-300115, Romania.*

**Appendix**

The electronegativity of an atom in a molecule ($\bar{\chi}$) can be evaluated starting from the atomic electronegativity ($\chi_0$) and hardness ($\eta_0$) through the relations [18a-c, 20]:

$$\overline{\chi} = \chi_0 - 2\eta_0 \Delta N$$

$$\overline{\chi} = \chi_0 \exp(-\gamma \Delta N) \cong \chi_0 - \chi_0 \gamma \Delta N \tag{A.1}$$

being $\gamma$ a fall-off parameter [18b].

The reliability of the proposed electronegativity expressions, eqs. 27 and 39, can be checked showing that they are consistent with A.1 equations.

To do this, one can consider the atom placed at the center of a sphere of infinite radius in which has to be evaluated the influence of the electronegativity of the same atom, up to where the probability to find electrons is very low. Therefore, if we apply the limit $N(r\to\infty)\to 0$, on the actual 27 and 39 absolute and Mulliken electronegativity formulations, the following equations are obtained:

$$\lim_{N(r\to\infty)\to 0} \chi(N) = -\frac{b}{a}$$

$$\lim_{N(r\to\infty)\to 0} \chi_M(N) = -\frac{b}{\sqrt{a}} \arctan\left(\frac{1}{\sqrt{a}}\right). \tag{A.2}$$

If the Poisson equation within the long-range condition is considered:

$$\nabla V(r) \cong -4\pi\rho(r)\Delta r,$$

$$V(r) \cong 4\pi\rho(r)[\Delta r]^2, \tag{A.3}$$

the components $a$ and $b$ in A.2 can be re-arranged, respectively, as:

$$a \cong \sum_{i=1}^{N} \frac{\nabla \rho_i}{4\pi \rho_i \Delta r_i} \Delta r_i = \frac{1}{4\pi}\sum_{i=1}^{N}\frac{\nabla \rho_i}{\rho_i} \cong \frac{1}{4\pi}\sum_{i=1}^{N}\frac{\Delta \rho_i}{\rho_i \Delta r_i} \cong \frac{1}{4\pi}\sum_{i=1}^{N}\frac{\Delta \rho_i}{\Delta \rho_i} = \frac{N}{4\pi},$$

$$b \cong \int \frac{\nabla \rho}{4\pi\rho\Delta r} 4\pi\rho[\Delta r]^2\, dr = \int \nabla\rho\, \Delta r\, dr \cong \int \frac{\Delta\rho}{\Delta r}\Delta r\, dr = \int \Delta\rho(r)\, dr = \Delta N. \tag{A.4}$$

Introducing A.4 in A.2, they become:

$$\lim_{r \to \infty} \chi(N) = -\frac{4\pi}{N}\Delta N \equiv \bar{\chi} - \chi_0$$

$$\lim_{r \to \infty} \chi_M(N) = -\sqrt{\frac{4\pi}{N}} \arctan\left(\sqrt{\frac{4\pi}{N}}\right)\Delta N \equiv \bar{\chi} - \chi_0 \tag{A.5}$$

that are formally identical to:

$$\bar{\chi} - \chi_0 = -2\eta_0 \Delta N \cong -\gamma\chi_0 \Delta N \ . \tag{A.6}$$